\documentclass[11pt]{article}
\usepackage{acl2015}
\usepackage{times}
\usepackage{url}
\usepackage{verbatimbox}
\usepackage{latexsym}
\usepackage{subcaption}
\usepackage{graphicx}
\usepackage{caption}
\graphicspath{ {files/} }
\usepackage{amsmath}
\usepackage{enumitem}

\title{Analyzing Network Effects on a Fanfiction Community}

\author{Andres Carvallo \\
   Pontificia Universidad \\Cat\'olica de Chile,\\ Santiago, Chile\\
   {\tt afcarvallo@uc.cl} 
  \\\And
  Denis Parra \\
   Pontificia Universidad \\Cat\'olica de Chile,\\ Santiago, Chile\\
  {\tt dparra@ing.puc.cl}  \\ }
  
\date{}

\begin{document}

\maketitle
\begin{abstract}
\small
Since the early days of the Web 2.0, online communities have been growing quickly and have become important part of life for large number of people. In one of these communities, fanfiction.net, users can read and write stories which are adapted, recreated and modified from original famous books, tv series, movies, among others. By following stories and their authors, the fanfiction community creates a social network. Previous research on online communities has shown how features of the social network can help explain the behavior of the community, so we are interested in studying fanfiction's social network as well as its influence in aspects of the community.  
In particular, in this article we describe several properties of the members of the community, and we also try to discover which factors explain the popularity of the authors. We discover that time since joining fanfiction and the size of the authors' biography, has a negative effect on the authors popularity. Moreover, we show that the users' network metrics help to explain better authors' popularity.


\end{abstract}

\section*{INTRODUCTION}

Online communities are one of the most important parts of the Web nowadays \cite{preece2005online}, tracing their roots to Internet news-based communities such as Usenet \cite{turner2005picturing}. There is a lack of consensus on the definition  of ``online community'', but in its most basic form, it has been defined
as ``software that allows people to interact and share
content in the same online environment'' \cite{malinen2015understanding}. A recent survey describes different types of online communities and it also identifies interesting research topics on this area. One of them, is whether or not it is possible
to identify other factors than the number of contributions supporting to the formation of thriving online communities  \cite{malinen2015understanding}. In this work we present preliminary results of a project aimed at identifying factors which explain success (or lack of it) in a large and well established online community for readers and writers, called fanfiction.net \footnote{https://www.fanfiction.net/}. In this community, users can read and write stories which are adapted, recreated and modified from original famous books, tv series, movies, among others. We present an analysis of a representative sample of user profiles and we study one aspect of this community: what makes authors more popular than others, with a particular focus on network features.

In the past, most of the work done on fanfiction.net is based on general data analysis such as sentiment analysis and community feedback \cite{evans2016more,milli2016beyond,yin2017no}, however, there has not been any work related to social network analysis. This paper attempts to provide a first sight to answering the following question: can social network features help to predict author's popularity? We explore at features like betweenness centrality, closeness centrality and pagerank. 

The paper is structured as follows. Section 2 provides the methodology used to solve the paper problem. Section 3 describes the data sample composed by user's profiles and reveals the distribution of favorite authors, favorite stories, stories written and communities. Then, in section 4 we construct the fanfiction network of favorite author relations. After that in section 5 we present the results of two logistic regression models to predict the popularity of an author using profile variables and network metrics. Finally in section 6 we discuss the results obtained and present the final conclusions.
\\

\subsection*{Related work}

Most of the studies of fanfiction.net have focused their analysis on how the community influences the quality of stories, such as the paper from Evans et al. \cite{evans2016more}, where they documented that as the story reviews increases, the authors receives feedback that allows them to somehow improve their writing quality and most of the times they change the story genre. Milli and Bamman \cite{milli2016beyond} have done computational analysis using language processing (NLP) to study features of the characters in the stories and compared the genre of written stories with the original work. They found that secondary characters tend to appear more in fanfiction stories than in the original works. Another interesting tendency found is the prevalence of female characters in stories. A great differentiator of this study is that they were the first to apply computational methods to analyze fanfiction data.

Another interesting work was done by Yin et al. \cite{yin2017no} that extracted a large amount of data made up from 6.8 million stories and 1.5 million authors in 44 different languages. They concluded that most of the stories are written in English, the Harry Potter saga is the franchise with more adapted stories, they analyzed the authors productivity over time, concluding that July is the month where most stories were published; they also discovered that the number of stories from 2013 to 2016 had increased considerably. Finally, they confirmed that the most frequent combination of genres in written stories are Romance and Humor, Anime is the most popular section and that books and games sections are more likely to receive more reviews.

The processes of studying the social network metrics an online community(Andreas Kaltenbrunner \textit{et al.} \cite{kaltenbrunner2011comparative}), estimation of factors to predict popularity of online news (Ksiazek et al. \cite{ksiazek2016user}), evaluation of network metrics on Wikipedia discussion pages (Laniado \textit{et al.} \cite{laniado2011wikipedians}) , interpretation of models to predict the popularity of online contents based on comments (Lee \textit{et al.} \cite{lee2010approach}), obtention of prediction values for contents future popularity (Ahmed \textit{et al.} \cite{ahmed2013peek} ), anticipation of popular twitter accounts (Imamori \textit{et al.} \cite{imamori2016predicting}) and prediction of posts popularity through comment mining (Jamali \textit{et. al} \cite{jamali2009digging}) has been a references for our study methodology. However in this work we do not focus on predicting the popularity of web contents or new user accounts. We concentrate on finding factors that affect the user's probability of having followers using network metrics and their profile information.

\subsection*{What is fanfiction?}
In this work we focus on data obtained from fanfiction.net, a long-established online community net, where users can share their stories, which are adapted, recreated and modified from original books, tv series, movies, video games, cartoons, among others. 
Registered users can: write stories, have favorite authors, have favorite other users written stories, add reviews to written stories and be part of an online community group.

The fanfiction.net online community consists of the following entities:

\begin{enumerate}

\begin{figure}[!htbp]
\centering
\includegraphics[width=77mm, height=70mm]{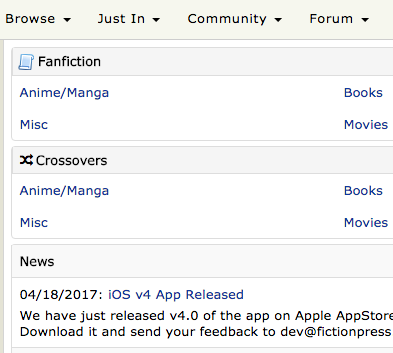}
\caption{\label{Fig. 1} fanfiction.net main page.}
\end{figure}

\item \textbf{Section}: the fanfiction online community homepage has three sections: fanfiction stories, crossovers and news. Fanfiction stories present different links for instance books, cartoons, movies and tv shows. Crossovers are stories that combine one or more franchises from multiple links, such as the book section, the movie section and vice versa. Finally the news section present updated headlines from the fanfiction community on a daily basis. (see Figure \ref{Fig. 1}) 

\begin{figure}[!htbp]
\small
\centering
\includegraphics[width=70mm]{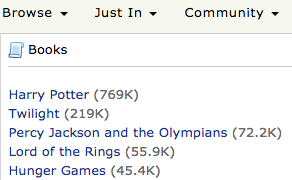}
\caption{\label{Fig. 2} fanfiction.net books section.}
\end{figure}

\item \textbf{Franchise Stories}: This section shows original pieces of work ranked according to the number of stories associated to the title.(see Figure \ref{Fig. 2}) 

\begin{figure}[!htbp]
\centering
\includegraphics[width=77mm, height=50mm]{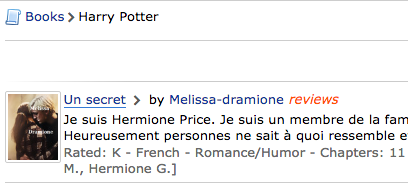}
\caption{\label{Fig. 3} fanfiction.net harry potter stories.}
\end{figure}

\item \textbf{Stories}: Once users have chosen their favorite franchise, they are presented with a list of fiction stories.These stories are recreated, adapted and modified from the original pieces of work written by other fanfiction members. Each story has a summary and links to the story itself, the authors profile and the story reviews.Every story box shows the rating, genre, language, chapters, words, favorites, follows and the story last time stamp update. (see Figure \ref{Fig. 3}) 

\begin{figure}[!htbp]
\centering
\includegraphics[width=70mm]{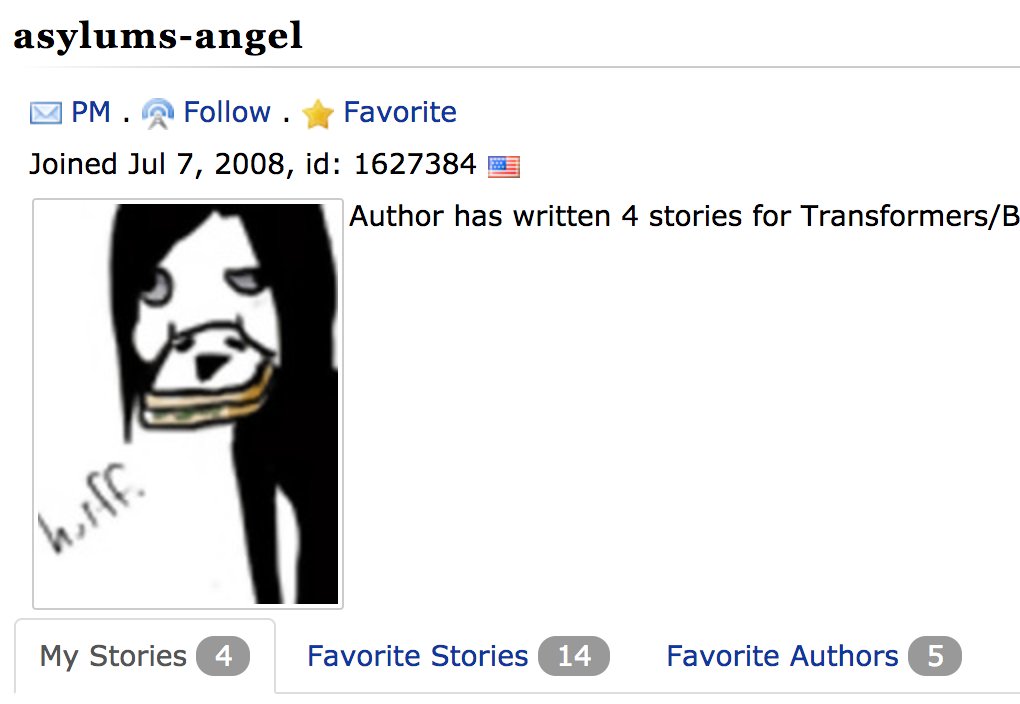}
\caption{\label{Fig. 4} fanfiction.net author profile.}
\end{figure}

\item \textbf{Authors}: They are fanfiction members who have written at least one adapted story from any original work. Their profile includes the signing up date, their user id, last update and biography. The author profile also shows their written stories, favorite stories, their favorite authors and the communities they belong to. 
(see Figure~\ref{Fig. 4}) 

\begin{figure}[!htbp]
\centering
\includegraphics[width=66mm, height=32mm]{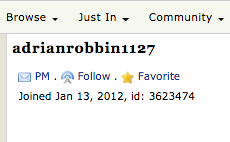}
\caption{\label{Fig. 5} fanfiction.net user profile.}
\end{figure}

\item \textbf{Users}: They interact with the community but do not have written stories. Users can have favorite stories, favorite authors and they can also review stories. Most of the times users present a profile that only includes their id, the signing up date and their last update.(see Figure~\ref{Fig. 5}) 

\item \textbf{Reviews}: They show the fanfiction feedback and comments given to the stories. This section also presents a list of reviews for every story, the time stamp and a link to the reviewers profile. (see Figure~\ref{Fig. 6}) 

\begin{figure}[!htbp]
\centering
\includegraphics[width=70mm, height=35mm]{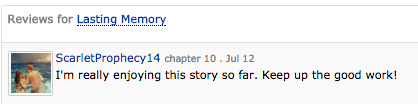}
\caption{\label{Fig. 6} fanfiction.net story reviews.}
\end{figure}

\item \textbf{Beta readers}: they are special users who monitor the community guidelines. They check the stories grammar, spelling, correctness and etiquette. Their profiles show their strengths, weaknesses, preferred language, genre and their favorite categories. (see Figure~\ref{Fig. 7})  

\begin{figure}[!htbp]
\centering
\includegraphics[width=65mm, height=33mm]{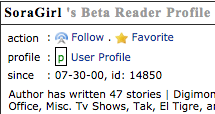}
\caption{\label{Fig. 7} fanfiction.net beta reader profile.}
\end{figure}

\end{enumerate}

\section*{METHODOLOGY}
To answer the paper question, if networks can help predicting authors popularity, we analyze the fanfiction.net social network metrics and contrast the results of two logistic regression models to predict popularity of an author, using the authors dataset. The first model considers only factors related to the authors profile such as: written stories, favorite stories, biography words, favorite authors and time spent since joining fanfiction. The second model, includes the variables of the first one plus users network metrics of betweenness centrality, closeness centrality, page rank and clustering coefficient.
\\

\section*{DATASET}

The dataset considers a sample of 83,600 users profiles from fanfiction.net over a total of 9,230,000 at 20th may 2017. The crawler of the data started on May 20th, 2017 and ended the 5th of June 2017. Almost 1,500 new users accounts are created every day, therefore any profile changes and deleted accounts after that period of time, are not considered for this study. 

From each user profile we extracted the following information: date they joined fanfiction, biography, written stories, favorite stories, favorite authors and communities if they are part of them.

\begin{table}[!htbp]
\small
\captionsetup{font=scriptsize}
\centering
\caption{Fanfiction users activity\\.}
\label{Table. 1}
\begin{tabular}{|c|c|c|c}
\hline 
\textbf{Users} & \textbf{Total} & \textbf{Percentage (\%)} \\ \hline
\begin{tabular}[c]{@{}c@{}}In community\\ No stories\\ No favorite stories\\No favorite authors\\ With no activity\\ \end{tabular} & \begin{tabular}[c]{@{}c@{}}1,175\\ 68,793\\ 56,015\\ 61,010\\ 49,422\\ \end{tabular} &\begin{tabular}[c]{@{}c@{}} 1,4\%\\ 82,2\%\\ 67,1\%\\ 72,9\%\\ 59,11\%\\ \end{tabular} \\ \hline
\end{tabular}
\end{table}

As seen in Table 1, most of fanction users have  no activity of writing, having favorite authors or favorite stories and/or being part of a community group. Users that are part of a community are only 1,175 (1.4 \% of the 83,599 users sample), users with no stories written are 68,793 (82,2 \% of the 83,599 users sample), users with no favorite stories 56,015 (67 \%), users with no favorite authors 61,010 (72,9\%) and finally, users with no activity in neither of those activities are close to the 50\% of the sample with 49,122 fanfiction users. 

Then in the following section we analyze distributions of users that are members a of a community group, have favorite stories, favorite authors and written stories.
\\

\begin{figure}[!htbp]
\centering
\includegraphics[width=90mm]{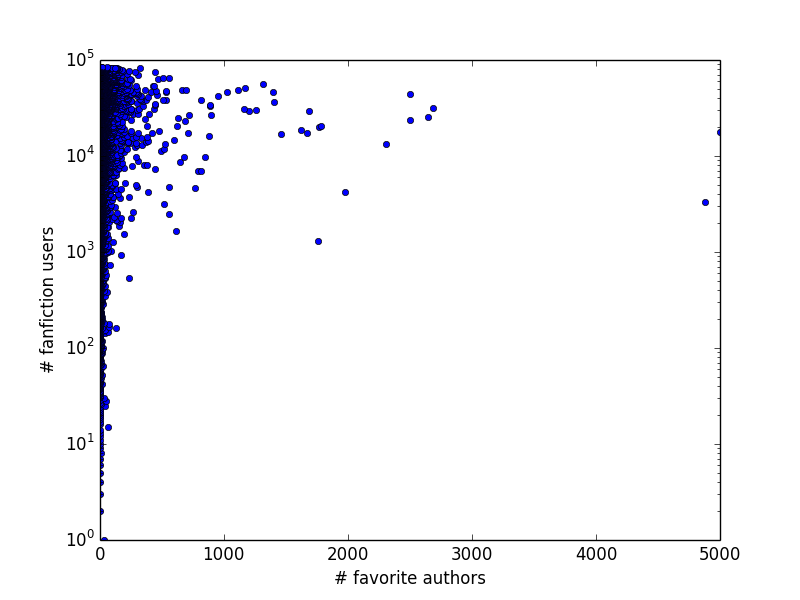}
\caption{\label{Fig. 8} Users with favorite authors distribution.}
\end{figure}

According to Figure~\ref{Fig. 8} that shows the distribution of users with favorite authors, it can be noticed that most of the users are distributed between zero and one, that means, that most of them do not have favorite authors, so in other words, they are passive members on the online community.  

\begin{figure}[!htbp]
\centering
\includegraphics[width=90mm]{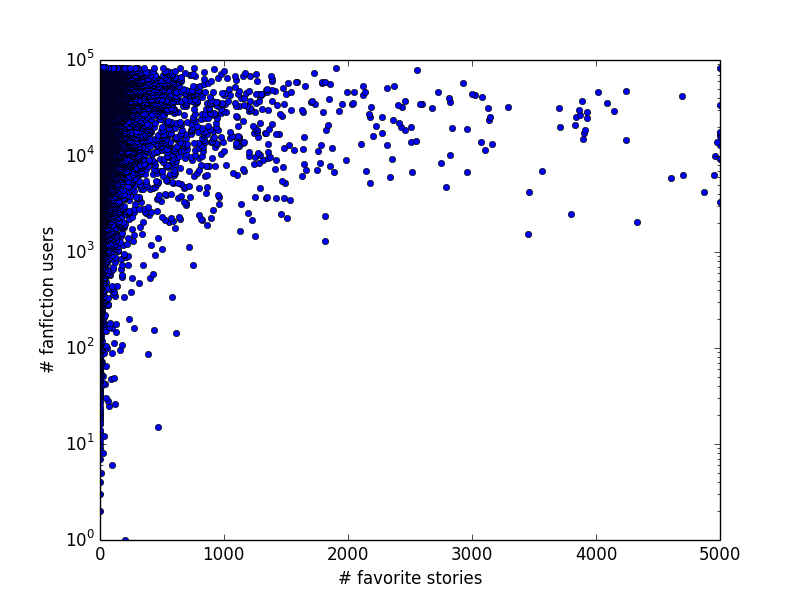}
\caption{\label{Fig. 9} Users with favorite stories distribution.}
\end{figure}

The same behavior is observed in Figure~\ref{Fig. 9}, which shows the distribution of users that have favorite stories, it presents that most of the users do not have favorite stories and that the users that have the higher amount of favorite stories account for approximately 350.

\begin{figure}[!htbp]
\centering
\includegraphics[width=82mm]{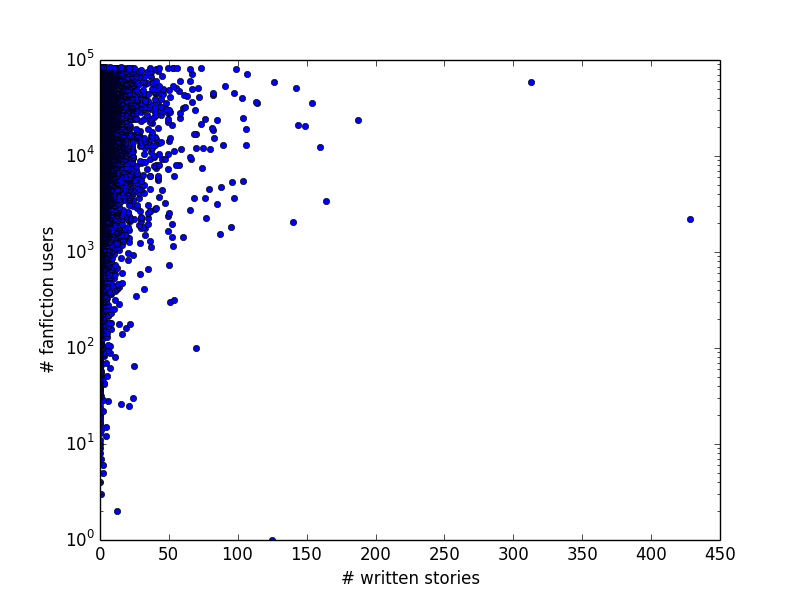}
\caption{\label{Fig. 10} Distribution of users with written stories.}
\end{figure}

Figure~\ref{Fig. 10} indicates the distribution of users who have written stories and it can be seen there is not so much activity in this item, which means they prefer being spectators than writing their own stories. 

\begin{figure}[!htbp]
\small
\captionsetup{font=scriptsize}
\centering
\includegraphics[width=70mm]{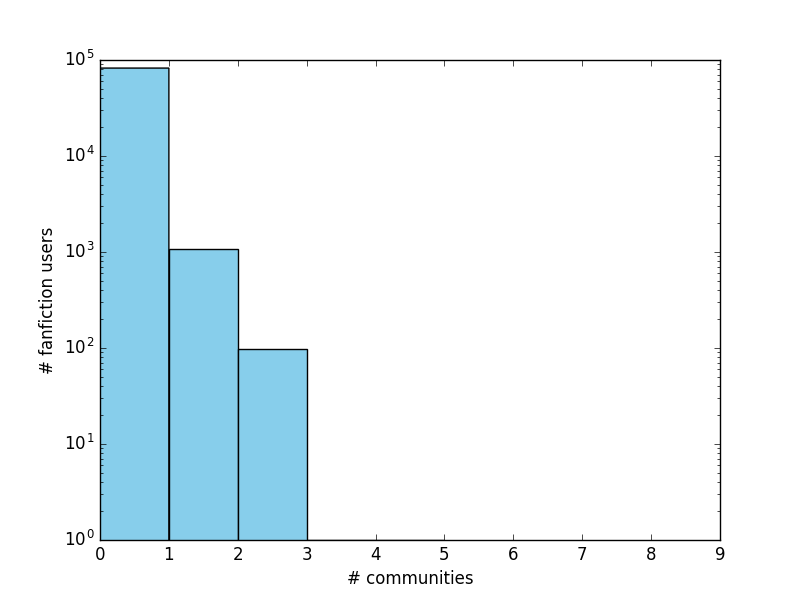}
\caption{\label{Fig. 11} Distribution of users in communities.}
\end{figure}

Users that belong to a community represent a small part of the sample. Figure~\ref{Fig. 11} shows the users in communities concentrate between zero and two communities per user. This tendency reveals that most of the times accepted users have to write stories.  
\\

\section*{ANALYSIS OF FANFICTION NETWORK}
In this section we construct the fanfiction.net network, which considers users that have at least one favorite author, where a relation between two nodes can be possible if one of them is the other's favorite author.

\begin{table}[!htbp]
\small
\captionsetup{font=scriptsize}
\centering
\caption{Global Measures of the fanfiction users favorite authors network\\.}
\label{Table. 1}
\begin{tabular}{|c|c|c|c}
\hline 
\textbf{Attribute} & \textbf{Result} \\ \hline
\begin{tabular}[c]{@{}c@{}}Average Degree\\ Network Diameter\\ Average Distance\\Number of Nodes\\ Number of Edges\\ \end{tabular} & \begin{tabular}[c]{@{}c@{}}854\\ 3\\ 1,112\\ 1,734\\ 2,766\\ \end{tabular}  \\ \hline
\end{tabular}
\end{table}

From the total sample of fanfiction users (83,650) we filtered the ones that had at least one favorite author, obtaining a total of 1,734 users and 2,766 interactions (as seen on Table 2).

The network is quite dispersed with an average degree of 854, an average diameter of 3 and an average distance of 1.12 between two nodes.

\begin{figure}[!htbp]
\captionsetup{font=scriptsize}
\centering
\includegraphics[width=60mm]{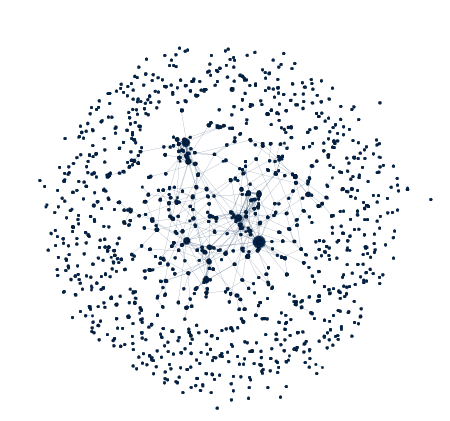}
\caption{\label{Fig. 12} Fanfiction users favorite authors network.}
\end{figure}

\begin{figure}[!htbp]
\captionsetup{font=scriptsize}
\centering
\includegraphics[width=60mm]{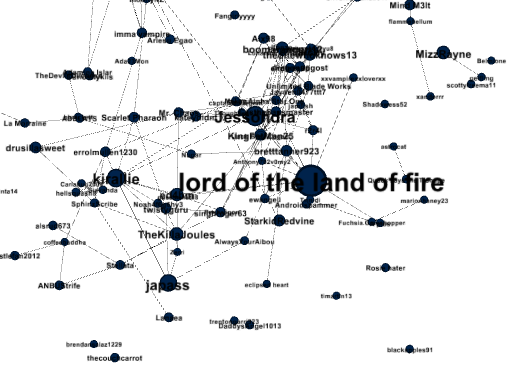}
\caption{\label{Fig. 13} Fanfiction users favorite authors node with bigger in and out degree.}
\end{figure}

For constructing the network (Figure~\ref{Fig. 12}) the size of the node depends on the quantity of in-degree relations. It can be observed that there is one node that concentrates most of the interactions and it is represented by the author "Lord of the Land Fire", which has the higher in degree indicator (see Figure~\ref{Fig. 13}). 

The network represents that most of the users do not have favorite authors and as seen in Figure ~\ref{Fig. 12} there is only one big cluster with a central node, which is the author with the highest in-degree indicator. 

Figure~\ref{Fig. 13} shows a close-up of the same network with authors names on the nodes. The bigger nodes represent the authors with the most in-degree relations, showing that "Lord of the Land of Fire" is the author with highest in-degree indicator and the most centered in the network, followed by the author "Jassondra". The values obtained from the in-degree indicators are shown in detail in the following tables. 

\begin{table}[!htbp] \centering
\captionsetup{font=scriptsize}
\tiny
  \caption{Betweenness Centrality} 
  \label{Table 2} 
\begin{tabular}{@{\extracolsep{1pt}}lcccc} 
\\[-1.8ex]\hline 
\hline \\[-1.8ex] 
Name & \multicolumn{1}{c}{\# stories}& \multicolumn{1}{c}{bfactor} & \multicolumn{1}{c}{category/franchises} & \multicolumn{1}{c}{\#dif franchises} \\ 
\hline \\[-1.8ex] 
Jessondra & 7 & 0.31 & Cartoons:Gargoyles & 1\\ 
Kiralie & 107 & 0.27 & Books:HarryPotter & 10\\ 
Lordlandfire & 85 & 0.23 & Anime:Youjo Senki & 15\\ 
Killajoules & 1 & 0.163 & Books: HarryPotter & 1\\ 
Guntz & 10 &  0.160 & Books: Hobbit & 7\\ 
\hline \\[-1.8ex] 
\end{tabular} 
\end{table} 

As shown in Table 3, the user with the highest betweenness centrality is "Jessondra" with a 0.31 indicator, presenting only 7 written stories and focused on a single cartoon franchise, Gargoyles. 

It is significant that, event though the author "Lord of the Land of Fire" represents the most centered node in the network figure, it is the author "Jessondra" the one who presents the highest betweenness centrality indicator. 

\begin{table}[!htbp] \centering
\captionsetup{font=scriptsize}
\tiny
  \caption{In-degree} 
  \label{Table 3} 
\begin{tabular}{@{\extracolsep{1pt}}lcccc} 
\\[-1.8ex]\hline 
\hline \\[-1.8ex] 
Name & \multicolumn{1}{c}{\# stories} & \multicolumn{1}{c}{in-degree} & \multicolumn{1}{c}{category/franchise} & \multicolumn{1}{c}{\#diff franchises}\\ 
\hline \\[-1.8ex] 
Lordofland & 85 & 76 & Anime: Youjo  Senki & 15  \\ 
Pattyrose & 18 & 46 & Books:Twilight & 1\\ 
Kirallie & 107 & 37 & Books:HarryPotter & 10\\ 
Halojones & 5 & 28 & Books:Twilight & 1\\ 
KingFatMan & 5 &  17 & Books:HarryPotter & 1\\ 
\hline \\[-1.8ex] 
\end{tabular} 
\end{table} 

One would tend to think that the more written stories, the greater the probability of having followers, but this is not the case, as seen in the in-degree ranking on Table 4, the user "Kirallie" has 107 written stories, which compared to the quantity of stories written by the author "Lord of the Land of Fire", the one who presents the highest in degree ranking. Nonetheless, the in degree indicator may be influenced by how the written stories are diversified within different franchises.

\begin{table}[!htbp] \centering
\captionsetup{font=scriptsize}
\tiny
  \caption{Page rank} 
  \label{Table 4} 
\begin{tabular}{@{\extracolsep{1pt}}lccccc} 
\\[-1.8ex]\hline 
\hline \\[-1.8ex] 
Name & \multicolumn{1}{c}{\# stories} & \multicolumn{1}{c}{pagerank} & \multicolumn{1}{c}{category/franchise} & \multicolumn{1}{c}{\# diff franchises}  \\ 
\hline \\[-1.8ex] 
Lordofland & 85 & 0.0213 & Anime: Youjo Senki & 15 \\ 
Pattyrose & 18 & 0.0115 &  Books:Hobbit & 7\\ 
Jassondra & 7 & 0.0098 & Cartoons: Gargoyles & 1\\ 
Kirallie & 107 & 0.0096 & Books: HarryPotter & 10\\ 
Halojones & 5 & 0.0065 & Books: Twilight & 1\\ 
\hline \\[-1.8ex] 
\end{tabular} 
\end{table}

The results obtained from the sample indicate that the "Lord of Land Fire" author is the one with the highest pagerank, as well as the author with the highest in degree index.

\begin{table}[!htbp] \centering
\captionsetup{font=scriptsize}
\tiny
  \caption{Clustering coefficient} 
  \label{Table 5} 
\begin{tabular}{@{\extracolsep{1pt}}lcccc} 
\\[-1.8ex]\hline 
\hline \\[-1.8ex] 
Name & \multicolumn{1}{c}{\# stories} & \multicolumn{1}{c}{factor} & \multicolumn{1}{c}{category/franchise} & \multicolumn{1}{c}{\# diff franchises}  \\ 
\hline \\[-1.8ex] 
Janedoe & 0 & 0.33 & - & - \\ 
Aplus & 16 & 0.06 & Books:HarryPotter & 1\\ 
r24l & 8 & 0.01 & Books:HarryPotter & 2\\ 
Jessondra & 7 & 0.001 & Cartoons: Gargoyles & 1\\ 
Lordofland & 85 & 0.0006 & Anime: Youjo Senki & 15\\ 
\hline \\[-1.8ex] 
\end{tabular} 
\end{table}

If we see the results from the clustering coefficient indicator ranking, which determines how nodes are connected to their neighbors, we can see that the author "Janedoe12345", has the highest clustering coefficient despite not being an author.

Table 6 indicates that the clustering coefficient ranking is segregated due to the differences between the author with the highest ranking and the others. The results show that the author "Aplus" has a five-time lower clustering coefficient indicator than "Janedoe". 

\begin{table}[!htbp] \centering
\captionsetup{font=scriptsize}
\tiny
  \caption{Closeness Centrality} 
  \label{Table 6 } 
\begin{tabular}{@{\extracolsep{1pt}}lcccc} 
\\[-1.8ex]\hline 
\hline \\[-1.8ex] 
Name & \multicolumn{1}{c}{\# stories} & \multicolumn{1}{c}{factor} & \multicolumn{1}{c}{category/franchise} & \multicolumn{1}{c}{\#diff franchises} \\ 
\hline \\[-1.8ex] 
Jessondra & 7 & 0.234 & Cartoons: Gargoyles & 1 \\ 
Kirallie & 107 & 0.231 & Books: HarryPotter & 10\\ 
Lordofland & 85 & 0.214 & Anime: Youjo Senki & 15\\ 
Guntz & 10 & 0.213 & Books: Hobbit & 7\\ 
Sml16 & 1 & 0.210 & Books: HarryPotter & 1\\ 
\hline \\[-1.8ex] 
\end{tabular} 
\end{table}

Table 7 points out that the author "Jassondra" has the highest closeness centrality indicator. Despite not having too many written stories compared to the author "Kirallie" and "Lord of the Land of Fire". The closeness centrality ranking presents an equitable distribution as it can be seen from the authors figures. 
\\

\section*{RESULTS}
Our next step is to study the factors that mostly affect the probability of an author to have an in-degree higher than one, for this analysis we use two logistic regression models on the authors dataset composed by 8,877 observations. The first model considers only factors related to the authors profile such as: written stories, favorite stories, biography words, favorite authors and time spent since joining fanfiction. The second model includes the variables of the first one plus users network metrics of betweenness centrality, closeness centrality, page rank and clustering coefficient. 

\begin{table}[htp] \centering 
\tiny
\captionsetup{font=scriptsize}
  \caption{Logit regression to predict the probability of an author of being followed considering network metrics and without network metrics} 
  \label{Table 7} 
\begin{tabular}{@{\extracolsep{3pt}}lcccc} 
\\[-1.8ex]\hline 
\\included &Model 1 & Model 2 \\ 
\hline \\[-1.8ex] 
 favorite\_authors & 0.091$^{**}$ & 0.068$^{*}$  \\ 
  & (0.0347) & (0.034) \\ 
  & \\ 
 favorite\_stories & 0.077$^{**}$ & 0.061$^{}$ \\ 
  & (0.028) & (0.036) \\ 
  & \\ 
 stories\_written & 0.88$^{***}$ & 0.848$^{***}$  \\ 
  & (0.043) & (0.047) \\ 
  & \\ 
 log\_time & $-$0.21$^{***}$ & -0.17$^{***}$ \\ 
  & (0.029) & (0.035)\\ 
  & \\ 
 log\_bio\_words & $-$0.017$^{}$ & -0.03$^{}$ \\ 
  & (0.023) & (0.028)\\ 
  & \\ 
 betweenness &  & -0.22$^{***}$ \\ 
  &  & (0.066)\\ 
  & \\ 
 clustering &  & 0.006$^{}$ \\ 
  &  & (0.119)\\ 
  & \\ 
 page rank &  & 0.2193$^{}$ \\ 
  &  & (0.1214)\\ 
  & \\ 
 closeness &  & 0.994$^{***}$ \\ 
  &  & (0.05)\\ 
  & \\ 
 Constant & $-$0.411$^{*}$ & -0.956$^{***}$ \\ 
  & (0.023) & (0.27)\\ 
  & \\  
observations & 8877 & 8877 \\ 
null deviance &  6890 & 6890 \\ 
residual deviance & 6071 & 4547 \\ 
AIC & 6083 & 4567 \\ 
Pseudo R2 (Mc Fadden) & 0.1188 & 0.34 \\

\hline 
\hline \\[-1.8ex] 
\textit{Note:}  & \multicolumn{1}{r}{$^{*}$p$<$0.1; $^{**}$p$<$0.05; $^{***}$p$<$0.01}  
\end{tabular} 
\end{table} 

It can be seen that both models share some significant variables such as number of favorite authors, stories written and time spent since joining to fanfiction. The number of written stories is the factor that mostly affects the probability of having followers, explained by its significance and the high coefficient value for both models. 

According to Table 8 a variable that has an unexpected behavior is time, which has a negative impact on the probability of having followers. Then, if we consider the network metrics variables from the second model, the clustering coefficient and the page rank are not significant. And despite of being significant, the betweenness centrality variable has an inverse relation to the probability of having followers.

Finally, regarding the model adjustment, the second one presents a significantly higher fitness, explained by lower AIC and lower deviance indicators. The pseudo r-squared is higher in the second model, with a value of 0.34. This means that, by including the network metrics, we can better explain the probability that an author has an in-degree higher than one in the fanfiction online community.
\\

\section*{DISCUSSION AND CONCLUSIONS}
In this study, we focus on answering the paper question, if networks can help predicting the authors popularity, to do that we have analyzed the fanfiction.net social network metrics and we obtained from the dataset sample that the network does not have the required density, we also discovered that there is a big cluster which was in the center of the network, and one author has concentrated most of the in-degree relations.
Regarding statistical analysis to answer the question, we have reached to interesting values, first and foremost, the model that considers network metrics allow us to better predict the probability to have potential followers in the future, comparing to the model that considers only information concerning to users. We also obtained that time spent in fanfiction has a negative effect on the dependent variable, which might be explained by the passivity of older users in online communities.

After obtaining the results our hypothesis was right within two of the important factors we thought were relevant for this study, that its to say, the closeness centrality and the quantity of written stories. On the other hand, the pagerank did not pay an important role when answering the paper question. 

In summary, the fanfiction.net network is not so dense because most of the users prefer reading other users stories to writing their own ones and they like better to have an empty profile than to interact with other users.  

For further studies, it might be interesting to identify how the probability of having an in-degree higher than one is affected by the users written stories reviews or also incorporate other variants as the story language, genre or country of origin of the writer.

\bibliographystyle{acl}
\bibliography{fanfiction}

\end{document}